\documentclass{sigchi-ext}
\usepackage[T1]{fontenc}
\usepackage{textcomp}
\usepackage[scaled=.92]{helvet} % for proper fonts
\usepackage{graphicx} % for EPS use the graphics package instead
\usepackage{balance}  % for useful for balancing the last columns
\usepackage{booktabs} % for pretty table rules
\usepackage{ccicons}  % for Creative Commons citation icons
\usepackage{ragged2e} % for tighter hyphenation

\def\plaintitle{Inclusivity and Surveillance} 

\def\emptyauthor{}
\def\plainkeywords{Inclusion; Decolonization; Anti-racism; Race positive design; }

\title{Challenges in Designing Racially Inclusive Language Technologies}

\numberofauthors{2}
\author{%
  \alignauthor{%
    \textbf{Kimi V. Wenzel}\\
    \affaddr{Carnegie Mellon University} \\
    \affaddr{5000 Forbes Ave, \\Pittsburgh PA 15213} \\
    \email{kwenzel@cs.cmu.edu} }\alignauthor{%
    \textbf{Geoff Kaufman}\\
    \affaddr{Carnegie Mellon University}\\
    \affaddr{5000 Forbes Ave, \\Pittsburgh PA 15213}\\
    \email{gfk@cs.cmu.edu} } }

\definecolor{linkColor}{RGB}{6,125,233}
\hypersetup{%
  pdftitle={\plaintitle},
  pdfauthor={\emptyauthor},
  pdfkeywords={\plainkeywords},
  bookmarksnumbered,
  pdfstartview={FitH},
  colorlinks,
  citecolor=black,
  filecolor=black,
  linkcolor=black,
  urlcolor=linkColor,
  breaklinks=true,
}

\begin{document}

%% For the camera ready, use the commands provided by the ACM in the Permission Release Form.
\CopyrightYear{2023}
\setcopyright{rightsretained}
\conferenceinfo{CHI'23}{April  23--28, 2023, Hamburg, Germany}
\isbn{xxx}
\doi{https://doi.org/10.1145/3334480.XXXXXXX}
%% Then override the default copyright message with the \acmcopyright command.
\copyrightinfo{\acmcopyright}

\maketitle

% Uncomment to disable hyphenation (not recommended)
% https://twitter.com/anjirokhan/status/546046683331973120
\RaggedRight{} 

% Do not change the page size or page settings.
\begin{abstract}
We take a critical lens toward the pursuit of racially inclusive language technologies and identify several areas necessitating future work. We discuss the potential harms of conversational technologies, outline three challenges that arise in inclusive design, and lastly, argue that conversational user interface designers and researchers should go beyond racially inclusive design toward anti-racist and race positive designs. %[Abstracts should be about 150]
\end{abstract}


% ACM Classfication


% Print the classficiation codes

% \marginpar{%
%   \vspace{-45pt} \fbox{%
%     \begin{minipage}{0.925\marginparwidth}
%       \textbf{Good Utilization of the Side Bar} \\
%       \vspace{1pc} \textbf{Preparation:} Do not change the margin
%       dimensions and do not flow the margin text to the
%       next page. \\
%       \vspace{1pc} \textbf{Materials:} The margin box must not intrude
%       or overflow into the header or the footer, or the gutter space
%       between the margin paragraph and the main left column. The text
%       in this text box should remain the same size as the body
%       text. Use the \texttt{{\textbackslash}vspace{}} command to set
%       the margin
%       note's position. \\
%       \vspace{1pc} \textbf{Images \& Figures:} Practically anything
%       can be put in the margin if it fits. Use the
%       \texttt{{\textbackslash}marginparwidth} constant to set the
%       width of the figure, table, minipage, or whatever you are trying
%       to fit in this skinny space.
%     \end{minipage}}\label{sec:sidebar} }

\section{Introduction}
Attention to ``ethical'' and ``inclusive'' developments in language technology has increased dramatically in recent years. A great deal of work has been devoted to understanding how racial bias can emerge in language model production, performance, and accuracy \cite{field2021survey}; however, work aimed at understanding how conversational user interfaces (CUIs) may themselves perpetuate racism is limited. In this paper, we introduce potential harms of racism in CUIs, discuss three broad challenges of inclusive design, and conclude by urging CUI researchers to invest in anti-racist and race positive design research. 

\section{Harms of Non-Inclusive CUIs}
Our recent work has revealed that miscommunication errors with conversational technologies can mirror that same effects as racist microaggressions in interpersonal interactions: A high speech recognition error rate can significantly increase Black users' levels of self-consciousness and significantly decrease their self-esteem and emotional affect. This effect was not found among white users \cite{wenzel2023}. Thus, CUIs that are not inclusive of marginalized dialects and vernaculars may pose not only a usability threat to these users but a psychological threat as well. These findings only increase the urgency for language technologists to create racially just CUIs.

\section{Three Challenges of Inclusive Design}
We outline three primary ethical challenges in designing racially inclusive conversational user interfaces: (1) Collecting representative voice and language data in a non-extractive, decolonial manner; (2) Understanding the CUI utility-surveillance trade off for vulnerable populations; and (3) Contributing to unmediated anti-racism.

%As Gangadharan and Niklas succinctly describe: ```Who' matters as much as 'how.''' 

%What remains constant is the marginality and deprivation experienced by socially silenced groups.

\subsection{Collecting representative voice and language data}
The high error rate in conversational technologies for marginalized users has largely been attributed to lack of representative voice and language data \cite{koenecke2020racial}. Thus, the direct solution to minimizing this error rate is to gather voice data from those underrepresented groups. The methods for achieving this solution, however, are conventionally colonial and harmful \cite{kouritzin2018toward, chilisa2019indigenous}. Many researchers have contributed frameworks and guidelines for non-extractive data collection \cite{igwe2022decolonising}; however, further investigation is needed to understand how these practices may uphold specifically in data collection for language models and conversational user interfaces. How might we create a representative language data set for conversational technologies while being respectful, non-extractive, and non-exploitative of the marginalized groups from whom we seek data from? Perhaps more importantly, in line with community-based participatory research practices \cite{hacker2013community}, we should not ask how we may seek data \textit{from} these groups, but rather, ask how we may seek data \textit{with} these groups. This further necessitates an understanding of whether or not these technologies are of significant value to marginalized groups at all. Even if such technologies appear to be of significance on a superficial level, utility-surveillance tradeoffs must also be considered and analyzed.

%Importantly, as Sloane et al. effectively state, we need to ``recognize design participation as work'' \cite{sloane2020participation}.

\subsection{Understanding surveillance risks}
%Even if we were able to create a representative voice data set, our issues with inclusion may not be fully resolved. 
Current research follows a rhetoric of techno-solutionism \cite{morozov2013save}, assuming that a positive outcome is inherent in racially inclusive technology. But how much would conversational technologies, even in the absence of their racial disparities, cause harm? There are various concerns regarding how emerging technologies can serve as tools to target and surveil marginalized groups \cite{gangadharan2017downside}. While there are many privacy researchers focusing specifically on conversational devices \cite{pfeifle2018alexa} and CUI researchers working on inclusive design \cite{sin2022alexa}, these two lines of work are often divorced. When designing CUIs for marginalized groups, how might we consider their unique privacy needs? In response, we urge the CUI community to consider increased collaborations with privacy and security researchers. 

\subsection{Contributing to unmediated anti-racism}
Framing the issue of inclusion as merely a dataset problem also ignores the structural and historical roots of discrimination, and fails to acknowledge the larger web of discrimination in which technology-mediated racism exists. Acknowledging this larger web of discrimination necessitates our understanding that CUIs are just one medium for present-day discrimination, and that mediums of discrimination have, and will continue to, change over time. While these mediums of discrimination may evolve, however, the targets of discrimination seldom do. Thus, we should not limit our focus solely to technology-mediated interactions. Prioritizing research on technology-mediated forms of discrimination often de-prioritizes unmediated forms of discrimination \cite{pena2019decentering}, ultimately implicitly suggesting that the \textit{form} of discrimination matters more than the marginalized groups and individuals themselves. To this end, we suggest CUI scholars and practitioners take care to reflect on how their work may contribute to anti-discrimination in unmediated contexts. This suggestion is in line with broader efforts to go beyond inclusivity, toward anti-racism.

\section{Beyond Inclusion: Toward Anti-Racism\\and Race Positive Design}
``Inclusion'' first and foremost centers the status-quo of privileged groups. ``Inclusion'' suggests that a privileged status-quo exists, and that previously excluded groups should be accepted to the said status-quo. Instead of simply aiming to offer marginalized groups the same opportunities as privileged groups through ``inclusion,'' we should (1) actively aim to dismantle the structures that caused this power imbalance through anti-racist practices and (2) center and celebrate silenced and oppressed identities through race positive design.

\subsection{Anti-racist design}
As Ibram X. Kendi writes, ``the only way to undo what is racist is to consistently identify and describe it -- and then dismantle it'' \cite{kendi2023antiracist}. Dismantling this structure requires comprehensive anti-racist education, and a reorienting of research and design work. How might we shift from critiquing, writing about, and discussing racism toward actively producing change in both our scholarly and local communities? (See \cite{abebe2022anti} and \cite{kendi2023antiracist} for further articulation and case studies.) 
%Anti-racism acknowledges that not only are equal opportunities important, but that \textit{equitable} outcomes are necessary \cite{berman2010racism}. By failing to achieve these equitable outcomes, inclusion often hides under the guise of ``not racist.'' However, in the fight against racism, non-racism does not exist. Non-racism is not the opposite of racism, anti-racism is. As Ibram X. Kendi writes, ``the only way to undo what is racist is to consistently identify and describe it -- and then dismantle it'' \cite{kendi2023antiracist}. Researchers may consider: xxxxx

%Linguistic scholar Motha writes, in response to the question: \emph{Is an antiracist and decolonizing applied linguistics possible?} "It is not inevitable. But it is possible. Every move we make, let us ask ourselves: Is it racist or anti-racist?, opening our eyes to the knowledge that there is no space of not-racist...'' \cite{motha2020antiracist}. %

\subsection{Race positive design}
Race positive design ``enables thinking about race as a positive presence—as cultural capital, histories of resistance, bindings between lands and peoples—and the means by which it can be a generative force in technologies for just and sustainable futures'' \cite{eglash2020race}. Recent work suggests that race positive identity affirmations may be a potential method for addressing the harms that discriminatory conversational technologies inflict on people of color; however, more work needs to be done to validate these design interventions \cite{wenzel2023}. Future work may consider: How might CUIs serve as a medium for race and culture to be celebrated? How might race and culture serve as a generative feature of language technologies, uplifting marginalized groups?

%As Anna Lauren Hoffman aptly puts: ``More inclusive datasets and data input fields billed as progress or `doing better' implicitly concede the harms inflicted by their less inclusive iterations, while celebrating narrow technical updates as a kind of victory for marginalized communities'' \cite{hoffmann2021terms}.

\section{Conclusion}
We outline three challenges in designing for inclusive conversational user interfaces, and provide suggestions for how researchers may begin to address these challenges: (1) Researchers should practice non-extractive decolonial data collection, and consider what legitimate value their data collection efforts may bring to the communities they are working with. (2) CUI designers and researchers should engage further in privacy and security work, especially as surveillance through CUIs may be especially harmful for marginalized populations. (3) Researchers should reflect on how their research on inclusivity in technology-mediated tools may contribute to broader unmediated anti-racism efforts. In addition to appreciating these challenges, we encourage researchers to consider race-positive CUI designs in their future work.

\balance{} 

\bibliographystyle{SIGCHI-Reference-Format}
\bibliography{sample}

\end{document}